\documentclass[sigconf,balance]{acmart}

\usepackage{latexsym}
\usepackage{graphicx}
\usepackage{subcaption}
\usepackage[T1]{fontenc}
\usepackage{xcolor}
\usepackage{listings}
\usepackage{amsbsy}
\usepackage{enumitem}

\definecolor{lightgray}{rgb}{.9,.9,.9}
\definecolor{darkgray}{rgb}{.4,.4,.4}
\definecolor{purple}{rgb}{0.65, 0.12, 0.82}
\definecolor{orange}{rgb}{0.91, 0.38, 0.09}
\definecolor{darkgreen}{rgb}{0.05, 0.46, 0.05}
\definecolor{darkred}{rgb}{0.576, 0.21, 0.203}
\definecolor{keycolor}{rgb}{0.00, 0.57, 0.99}
\definecolor{codegreen}{rgb}{0.31, 0.59, 0.03}
\definecolor{codegray}{rgb}{0.5,0.5,0.5}
\definecolor{codepurple}{rgb}{0.58,0,0.82}
\definecolor{codecomment}{rgb}{.6,.6,.6}

\lstdefinestyle{PythonStyle}{
    commentstyle=\color{codecomment},
    emph={pandas, DataFrame, asapy, ExecutionController},
    emphstyle=\color{codegreen},
    keywordstyle=\ttfamily\bfseries\color{keycolor},
    numberstyle=\color{codegray},
    stringstyle=\color{codepurple},
}

\newif\ifcolonfoundonthisline

\makeatletter

\lstdefinelanguage{JsonPlus}
{
  showstringspaces    = false,
  keywords            = {false,true},
  commentstyle        = \color{purple},
  alsoletter          = 0123456789.,
  morestring          = [s]{"}{"},
  stringstyle         = \ifcolonfoundonthisline\color{orange}\else\color{keycolor}\fi,
  MoreSelectCharTable =%
    \lst@DefSaveDef{`:}\colon@json{\processColon@json},
  basicstyle          = \footnotesize,
  keywordstyle        = \ttfamily\bfseries,
}

\newcommand\processColon@json{%
  \colon@json%
  \ifnum\lst@mode=\lst@Pmode%
    \global\colonfoundonthislinetrue%
  \fi
}

\lst@AddToHook{Output}{%
  \ifcolonfoundonthisline%
    \ifnum\lst@mode=\lst@Pmode%
      \def\lst@thestyle{\color{darkred}}%
    \fi
  \fi
  \lsthk@DetectKeywords%
}

\lst@AddToHook{EOL}%
  {\global\colonfoundonthislinefalse}

\newcommand\YAMLcolonstyle{\color{red}\mdseries}
\newcommand\YAMLkeystyle{\color{black}\bfseries}
\newcommand\YAMLvaluestyle{\color{blue}\mdseries}

\makeatletter

\newcommand\language@yaml{yaml}

\expandafter\expandafter\expandafter\lstdefinelanguage
\expandafter{\language@yaml}
{
  keywords={true,false,null,y,n},
  showspaces=false,
  keywordstyle=\color{darkgray}\bfseries,
  basicstyle=\YAMLkeystyle,                                 
  sensitive=false,
  comment=[l]{\#},
  morecomment=[s]{/*}{*/},
  commentstyle=\color{purple}\ttfamily,
  stringstyle=\YAMLvaluestyle\ttfamily,
  moredelim=[l][\color{orange}]{\&},
  moredelim=[l][\color{magenta}]{*},
  moredelim=**[il][\YAMLcolonstyle{:}\YAMLvaluestyle]{:},   
  morestring=[b]',
  morestring=[b]",
  literate =    {---}{{\ProcessThreeDashes}}3
                {>}{{\textcolor{red}\textgreater}}1     
                {|}{{\textcolor{red}\textbar}}1 
                {\ -\ }{{\mdseries\ -\ }}3,
}

\lst@AddToHook{EveryLine}{\ifx\lst@language\language@yaml\YAMLkeystyle\fi}
\makeatother

\newcommand\ProcessThreeDashes{\llap{\color{cyan}\mdseries-{-}-}}

\usepackage{svg}

\copyrightyear{2024}
\acmYear{2024}
\setcopyright{acmlicensed}\acmConference[SIGSIM PADS '24]{38th ACM SIGSIM Conference on Principles of Advanced Discrete Simulation}{June 24--26, 2024}{Atlanta, GA, USA}
\acmBooktitle{38th ACM SIGSIM Conference on Principles of Advanced Discrete Simulation (SIGSIM PADS '24), June 24--26, 2024, Atlanta, GA, USA}
\acmDOI{10.1145/3615979.3656063}
\acmISBN{979-8-4007-0363-8/24/06}

\begin{document}

\title{AsaPy: A Python Library for Aerospace Simulation Analysis}

\author{Joao~P.~A.~Dantas}
\email{dantasjpad@fab.mil.br}
\orcid{0000-0003-0300-8027} 
\affiliation{%
  \institution{Institute for Advanced Studies}
  \city{Sao Jose dos Campos}
  \country{Brazil}
}

\author{Samara~R.~Silva}
\email{samarasrs@fab.mil.br}
\orcid{0009-0000-5065-7236} 
\affiliation{%
  \institution{Institute for Advanced Studies}
  \city{Sao Jose dos Campos}
  \country{Brazil}
}

\author{Vitor~C.~F.~Gomes}
\email{vitorvcfg@fab.mil.br}
\orcid{0000-0003-3239-2160} 
\affiliation{%
  \institution{Institute for Advanced Studies}
  \city{Sao Jose dos Campos}
  \country{Brazil}
}

\author{Andre~N.~Costa}
\email{negraoanc@fab.mil.br}
\orcid{0000-0002-2309-9248} 
\affiliation{%
  \institution{Institute for Advanced Studies}
  \city{Sao Jose dos Campos}
  \country{Brazil}
}

\author{Adrisson~R.~Samersla}
\email{samerslaars@fab.mil.br}
\orcid{0009-0005-4559-1849} 
\affiliation{%
  \institution{Institute for Advanced Studies}
  \city{Sao Jose dos Campos}
  \country{Brazil}
}

\author{Diego~Geraldo}
\email{diegodg@fab.mil.br}
\orcid{0000-0003-1389-9142} 
\affiliation{%
  \institution{Institute for Advanced Studies}
  \city{Sao Jose dos Campos}
  \country{Brazil}
}

\author{Marcos~R.~O.~A.~Maximo}
\email{mmaximo@ita.br}
\orcid{0000-0003-2944-4476} 
\affiliation{%
  \institution{Aeronautics Institute of Technology}
  \city{Sao Jose dos Campos}
  \country{Brazil}
}

\author{Takashi~Yoneyama}
\email{takashi@ita.br}
\orcid{0000-0001-5375-1076} 
\affiliation{%
  \institution{Aeronautics Institute of Technology}
  \city{Sao Jose dos Campos}
  \country{Brazil}
}

\renewcommand{\shortauthors}{Joao~P.~A.~Dantas et al.}

\begin{abstract}
AsaPy is a custom-made Python library designed to simplify and optimize the analysis of aerospace simulation data. Instead of introducing new methodologies, it excels in combining various established techniques, creating a unified, specialized platform. It offers a range of features, including the design of experiment methods, statistical analysis techniques, machine learning algorithms, and data visualization tools. AsaPy's flexibility and customizability make it a viable solution for engineers and researchers who need to quickly gain insights into aerospace simulations. AsaPy is built on top of popular scientific computing libraries, ensuring high performance and scalability. In this work, we provide an overview of the key features and capabilities of AsaPy, followed by an exposition of its architecture and demonstrations of its effectiveness through some use cases applied in military operational simulations. We also evaluate how other simulation tools deal with data science, highlighting AsaPy's strengths and advantages. Finally, we discuss potential use cases and applications of AsaPy and outline future directions for the development and improvement of the library.
\end{abstract}

\ccsdesc[500]{Information systems~Data analytics}
\ccsdesc[500]{Mathematics of computing~Statistical paradigms}
\ccsdesc[500]{Applied computing~Aerospace}
\ccsdesc[500]{Applied computing~Military}

\keywords{Aerospace Simulations, Data Analysis, Design of Experiments, Machine Learning, Military Scenarios.}

\maketitle

\section{INTRODUCTION}
\label{sec:introduction}

The application of simulation technologies in aerospace has significantly expanded, notably in commercial aviation, space exploration, and particularly in the military realm~\cite{costa2021formation}. This shift from live exercises to simulation is due to multiple reasons, including cost reduction and increased safety~\cite{davis2007analysis}. Simulation may be used for designing, testing, and optimizing complex systems such as aircraft, radars, and weapons~\cite{dantas2021weapon}. However, the vast amount of simulation data generated can be overwhelming, making the analysis process time-consuming and challenging, which may require more sophisticated algorithms and tools~\cite{dantas2022supervised,dantas2022machine}. In response, AsaPy, a custom-made Python library, was designed in the context of the Aerospace Simulation Environment (\emph{Ambiente de Simulação Aeroespacial -- ASA} in Portuguese)~\cite{dantas2022asa, dantas2023asasimaas} to simplify and expedite the analysis of military simulation data to support the decision-making process.

Rather than introducing new methods, AsaPy specializes in integrating a range of established techniques into a cohesive and specialized toolkit, adept at meeting the complex needs of aerospace data analysis. AsaPy offers a comprehensive pipeline of routines that typically would be performed step by step by researchers, including pre-checks before employing a specific analysis method, for example. Integrating processes into a single workflow makes AsaPy accessible even to those not proficient in programming, enabling them to apply robust analysis to aerospace data. The library includes features such as experimental design methods, statistical analysis, machine learning algorithms, and data visualization tools. This array of tools allows engineers and researchers to extract valuable insights from aerospace simulations, applicable not just in military scenarios but also in civilian and commercial aerospace sectors. Initially developed to operate alongside ASA, AsaPy's adaptable architecture also supports its use with other simulation frameworks, as will be demonstrated in the use cases section, evidenced by its integration of recognized scientific computing libraries like NumPy~\cite{harris2020array}, SciPy~\cite{virtanen2020scipy}, and Scikit-learn~\cite{pedregosa2011scikit}, ensuring both high performance and scalability.

The main contribution of this work is to provide an overview of the key features and capabilities of AsaPy, including its structure, effectiveness, and potential applications, mainly for analyzing aerospace and military simulation data. We also review some of the available simulation software, focusing on what data science capabilities they provide. Additionally, we bring some use cases to the AsaPy library applied to the defense context. Finally, we outline future directions for the development and improvement of the library.  We have provided a link to AsaPy and hope that this library proves beneficial for other analysis projects.

\section{RELATED WORK}
\label{sec:related-work}
Being conceived as a part of the ASA suite, AsaPy provides an integrated solution for data science activities within a Computer-Generated Forces (CGF) package. In this context, we focused on evaluating existing CGF tools with respect to their data science features. This evaluation was made following a similar methodology as seen in~\citet{abdellaoui2009comparative} and~\citet{toubman2015modeling}. Additionally, we provide an overview of the research background on data farming and Knowledge Discovery in Simulation data (KDS), fundamental concepts for the development of AsaPy.

\subsection{Existing Solutions}

\citet{abdellaoui2009comparative}~conducted a similar analysis and comparison of various modeling and simulation packages, with particular emphasis on their artificial intelligence (AI) capabilities. The evaluation was based on five crucial factors: architecture, autonomous operation, learning, organization, and realism. Despite the comprehensive nature of the study, notice that the authors only briefly touched on the role of data science in the context of these packages. Specifically, they made a passing reference to the existence of entity databases without delving into how data science principles might be applied to analyze simulation results. 

\citet{toubman2015modeling}~specifically examined the computer-generated forces (CGF) learning capabilities. Although they suggested using data for machine learning algorithms to extract behavior rules and apply them to new situations, their study did not address how commercial off-the-shelf (COTS) and government off-the-shelf (GOTS) products handle this approach. Moreover, they did not discuss how to analyze simulation data to derive general conclusions from scenario results.

We aim to expand these analyses by evaluating the status of data science capabilities within simulation packages, which would benefit researchers and practitioners in this field. Therefore, we conducted a survey of publicly available product information for the same COTS products (GOTS were excluded since they are not internationally available) as listed in Table~\ref{tabProducts}, each of them briefly described as follows.

\begin{table}[htb]
\small
\centering
\caption{Mention of “Data Analysis” and “Design of Experiments” (DoE), on the websites of seven COTS CGF packages (in no particular order).}
\label{tabProducts}
\begin{tabular}{llcc}
\hline
\vtop{\hbox{\strut Product}\hbox{\strut Name}} & \vtop{\hbox{\strut Company}\hbox{\strut Name}} & \vtop{\hbox{\strut Mention of}\hbox{\strut Data Analysis}} & \vtop{\hbox{\strut Mention of}\hbox{\strut DoE}} \\ 
\hline
STAGE            & Presagis            & No  & No \\
VR-Forces        & MAK Technologies    & No  & No \\
SWORD            & MASA                & No  & No \\
VBS4             & Bohemia Interactive & No  & No \\
DirectCGF        & Diginext            & No  & No \\
Steel Beasts Pro & eSim Games          & No  & No \\
FLAMES           & Ternion             & Yes & Yes \\
\hline
\end{tabular}
\end{table}

Scenario Toolkit and Generation Environment (STAGE)~\cite{stage} is a stand-alone synthetic tactical simulation software that facilitates the development of models for complex war scenarios involving various platforms such as avionics, naval, and land systems. The software comes equipped with models of multiple sensors, including radar, sonar, and missile warning systems, as well as weapons such as missiles and guns~\cite{anghinolfi2013agent}.

VR-Forces~\cite{vrforces} is a simulation environment that enables the generation of multiple scenarios. The software is equipped with features required for use as a tactical leadership trainer, a threat generator, a behavior model test bed, or a Computer Generated Forces (CGF) application~\cite{pullen2012open}.

SWORD~\cite{masasword} is a software suite comprising scenario creation applications, aggregated constructive simulation, and analysis tools specifically designed for staff training, education, classroom teaching, planning support, analysis, operational research, and C2 system stimulation. It is a comprehensive solution that allows simulation of operations ranging from battalion to division level and is the leading software provider in the market for training tactical level land staffs~\cite{drozd2023effectiveness}.

Virtual Battlespace 4 (VBS4)~\cite{vbs4} is a virtual and constructive simulation platform that enables the creation and execution of military training scenarios. Its workflow and features allow for quick training initiation, simplified editing and updating of training scenarios and terrains, and collaborative training simulations across any location on its virtual Earth~\cite{evensen2017modelling}.

DirectCGF~\cite{directCGF} is a battlespace generation software by DIGINEXT, which utilizes the simulation engine DirectSim. It comes equipped with a collection of pre-built models such as platforms, sensors, weapons, electronic warfare, and communication, along with automatic and intelligent behavior. Its modular architecture facilitates reusability and enhances productivity gains, as users can integrate dedicated plug-ins into the system~\cite{treatymodelling}.

Steel Beasts~\cite{steelBeasts} is a simulation tool that models armored warfare scenarios. Military forces around the world use it to support training, mission rehearsal, and analysis of vehicle-centered scenarios featuring gunners, commanders, and drivers~\cite{mckeown2012correlated}.

FLAMES~\cite{flames} is a range of products that offer a framework for custom constructive simulations designed to meet the specific requirements of the aerospace, defense, and transportation industries. It includes customizable scenario creation, execution, visualization, and analysis, as well as interfaces to constructive, virtual, and live systems~\cite{oeztuerk2016grundsatzbetrachtung}. It is the only one of the reviewed COTS packages that explicitly addressed the DoE and data analysis aspects, providing some options in its enhanced analysis option. Regarding the DoE, it accepts a manual setup in tables, as well as importing experiment files defined by third-party tools. With respect to data analysis, it also mentions third-party programs that can deal with user-specified output files.

Besides these COTS products, we would like to mention a GOTS package that is particularly relevant to the context of this paper. The Advanced Framework for Simulation, Integration and Modeling (AFSIM) is an objected-oriented C++ library used to create simulations in aerospace and defense contexts. It provides a range of features for simulating and analyzing complex operational scenarios, including air-to-air combat, air-to-ground strike, and reconnaissance missions~\cite{clive2015advanced}. With the focus on data science capabilities, we can point out the Visual Environment for Scenario Preparation and Analysis (VESPA), which supports creating scenario initial condition files that are compatible with AFSIM-based applications, enabling its usage as a DoE tool.

In summary, all the reviewed solutions lack an integrated approach with more comprehensive data science tools. FLAMES and AFSIM seem to be the closest to what AsaPy aims to provide within the context of ASA. However, they still rely on third-party packages and focus on data recording and visualization rather than the analysis itself.

\subsection{Related Concepts}

Originally, the concept of data farming emerged in the context of military simulations, providing decision-makers with the ``Commander’s Overview'' for enhanced decision support~\cite{horne2014data}. By encompassing a broad spectrum of parameter spaces and having the ability to explore both positive and negative effects, relationships, and potential options, data farming may unveil aspects not previously addressed in military simulation applications.

In other words, data farming has been employed to describe the intentional generation of data from simulation models. Through extensive designed experiments, one can efficiently and effectively ``cultivate'' simulation output. This approach allows for exploring vast input spaces and discovering noteworthy features in complex simulation response surfaces. Embracing this innovative mindset enables significant advancements in the scope, depth, and timely acquisition of insights provided by simulation models~\cite{lucas2015changing}.

There are three primary goals in data farming~\cite{kleijnen2005user}: (i) developing a fundamental understanding of the simulation model and the emulated system; (ii) identifying robust policies and decisions; and (iii) comparing the merits of various policies or decisions. Well-designed experiments prove to be efficient and effective tools for achieving these objectives. Despite the exponential increase in processing capabilities, the strategic design of experiments remains essential for obtaining comprehensive insights through large-scale simulation studies.

The KDS combines data farming with visual analytics-based methods~\cite{feldkamp2020knowledge}. The idea is to start by defining experiments, focusing on the selection of factors. Factors can encompass a wide array, including structural, organizational, technical data, system load, and material flow data. The number of factors, along with the lower and upper-value limits for each factor, is chosen as expansively as possible unless they are evidently irrelevant or physically implausible. For data generation, the simulation model is treated as a black box that transforms a set of input factor values into output data. This output data is then stored in a simulation output database, where a row of parameter values represents each experiment. Notably, experiments can be easily distributed across parallel machines as they are independent of each other. Once all experiments are conducted, the data analysis phase can proceed. The initial analysis begins with the simulation output data, and subsequently, knowledge is derived, especially when exploring relationships between output data and input factors. Well-suited visualizations play a crucial role in establishing connections between corresponding input and output sets, enabling users to investigate and draw conclusions.

Both data farming and its combination with visual analytics (KDS) have been the basis for establishing AsaPy features, which aim to provide means for not only intentionally generating simulation data but also obtaining insights from this data.

\section{STRUCTURE}
\label{sec:architecture}


From design, we aimed to develop a library that would help analyze simulation data, especially for military scenarios. In the ASA context, we noted that the analyst tends to follow a pattern that can be broken into four steps, shortly summarized: 

\begin{enumerate}
    \item Design of Experiments, in which we define the input configuration for the executions;
    \item Execution Control, in which we monitor the progress of a batch of executions;
    \item Analysis, in which we conduct the actual data analysis; and
    \item Prediction, in which we train a model to predict the outcome of new input configurations. 
\end{enumerate}

The literature concerning data analysis is vast, therefore it is important for our architecture not to limit the options available to the analyst. For this reason, we opted to package existing Python libraries that implement the desired methods, leveraging the benefits of this ecosystem and allowing easy interoperability. 

Therefore, our structure consists of a curated set of third-party libraries wrapped in a standardized and extensible way. The code is divided into four modules: \texttt{analysis}, \texttt{models}, \texttt{doe}, and \texttt{utils}, which roughly resembles the steps mentioned above. 

\begin{figure*}[ht]
    \centering
    \includesvg[width =\textwidth]{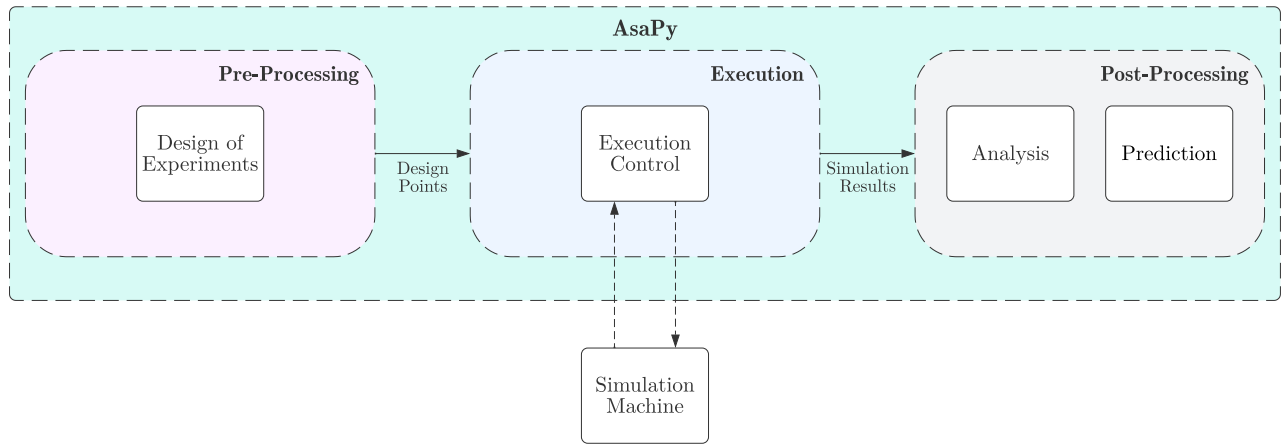}
    \caption{The structure of the AsaPy library in the perspective of the analyst workflow.}
    \label{fig:structure}
\end{figure*}

With this architecture, illustrated in Figure~\ref{fig:structure}, we allow the analysts to use the techniques independently or in combination or even to extend the package with other desired methods. Furthermore, we can automatize the process for the analyst, choosing reasonably appropriate tools for specific tasks, thus guiding his work to conduct a proper analysis.  


Although designed to be used integrated with other ASA suite tools, AsaPy is generic with respect to the simulation machine chosen to actually execute the scenarios and generate the data, as discussed in Subsection~\ref{subsec:execution_control}. For illustration, the integration with ASA service for parallel batch execution is implemented in the \texttt{asa-client} package, which has features such as authentication on the ASA platform, access to available scenarios, submission of experiment executions, and retrieval of execution results, not the focus of this work though.

In the following subsections, we discuss the techniques available for each one of the mentioned steps, locating them in the library structure.  

\subsection{Design of Experiments}

The Design of Experiments (DoE) step, enabled through the \texttt{doe} module, integrates a comprehensive array of tools and features for experimental design, as outlined in \citet{montgomery2017design}. Presently, the module offers the Latin Hypercube Sampling (LHS) technique~\cite{box2005statistics} as its primary method. Plans to augment this suite with advanced methods are underway, with particular emphasis on incorporating the Nearly Orthogonal Latin Hypercube (NOLH) technique, a more refined approach discussed in detail by \citet{cioppa2007efficient}.

This module is adept at managing various data types, including numerical, categorical, and boolean. It is instrumental in generating input samples for diverse simulation executions. Typically, these simulations run in batch mode, subject to fluctuating input parameters. The module also includes strategies for orchestrating metrics to assess simulation performance. This aspect is crucial for making informed decisions about whether to prematurely halt batch execution, a topic further explored in Subsection~\ref{subsec:execution_control}.

To utilize this tool effectively, analysts must specify which input variables are subject to modification, either manually or via automation (as implemented in the \texttt{asa-client} using the ASA suite). Following selecting an appropriate sampling technique, it generates the configurations for execution, thereby channeling design points into the subsequent phase of the process.

\subsection{Execution Control}
\label{subsec:execution_control}

Once the scenarios are created and the input parameters assigned, the next step is to run such experiments. Therefore, in this subsection, we discuss the control of the executions, including the process of splitting the total amount of runs into chunks and evaluating metrics to determine whether to stop the batch execution early. 

In the military context, the desired analysis is usually complex, requiring many executions to extract meaningful information. For this reason, the individual runs are usually dispatched as a batch, optimizing the usage of the computational resources available. However, often, not all executions planned are necessary for the analysis, which unfortunately can only be known during its execution. The large batch is broken into chunks of experiments to handle this limitation. Then, each chunk is sequentially executed until completion using all the available computational resources. After each chunk completion, the concerned metrics can be analyzed to decide whether to stop the batch execution or start the next chunk. To do so, one can observe the variation of the expected value of a significant variable before and after running the last chunk. If it is below a certain threshold, we can assume that this statistic has converged, economizing on the number of individual executions. 

Though naive, this heuristic, actually implemented by AsaPy, demonstrates how to apply a method to early stop a batch execution, saving time for the analyst and reducing the usage of the computational resources. Naturally, these evaluation metrics used for early stop criteria will depend on the objectives of the simulation. 

For instance, consider a scenario in the defense context where simulations are conducted in batches to optimize the number of aircraft needed to neutralize all enemy aircraft. Evaluation metrics for each simulation might encompass the number of remaining enemy aircraft and the number of missiles expended by the conclusion of each simulation. The unique aspect here is the introduction of early stop criteria, which are assessed not for individual simulations but across the entire batch. Should a significant portion of simulations within the batch meet these criteria early on, the whole batch can be terminated beforehand. Results from the simulations completed up to that point are then analyzed. This method proves efficient, conserving both time and resources, especially when the criteria are satisfied early in the batch run. It is important to note that the end of an individual simulation is controlled by the parameters set in the simulation file and is not directly associated with AsaPy.

Describing its intended usage, the control of executions is carried out by the \texttt{ExecutionController} class, which is instantiated with two functions and one number, as exemplified in Listing~\ref{lst:ec_example}. The first argument is a function that is responsible for effectively running the executions: receiving the collection of design points and returning the corresponding results. Moreover, the second function is the actual stop criteria, as already mentioned. Finally, the third argument is the size of the chunks into which the batch will be split. \\


\begin{lstlisting}[label={lst:ec_example},caption={Usage example of the Execution Control module}, language=Python, style=PythonStyle]
def simulate(doe: pandas.DataFrame) -> pandas.DataFrame:
    # 1. send execution requests using the Asa-client
    # 2. retrieve executions results
    return pandas.DataFrame.from_dict(asa_results)

def stop_check(result: pandas.DataFrame, last_result: pandas.DataFrame) -> bool:
    # compare results using Asapy or custom functions
    return compare_results(result, last_result)

ec = asapy.ExecutionController(simulate, stop_check, 100)
result = ec.run(doe)
\end{lstlisting}
 
Notice that this functional style allows the library client code to define how the simulations should be executed and what criteria should be used. This empowers analysts to use whichever simulation machine they may desire, just requiring the implementation of a function that interfaces with the chosen simulator and respects the expected signature. This interoperability easiness was one of the library design cardinal points and is now illustrated. 

\subsection{Analysis}

The analysis step is supported by the \texttt{analysis} module and provides a range of tools for analyzing and exploring simulation data. Therefore, in this section, we discuss some of the available components and how the package automatically chooses an appropriate technique for the analyst.

\begin{figure}[ht]
\centering
    \includesvg[width=0.485\textwidth]{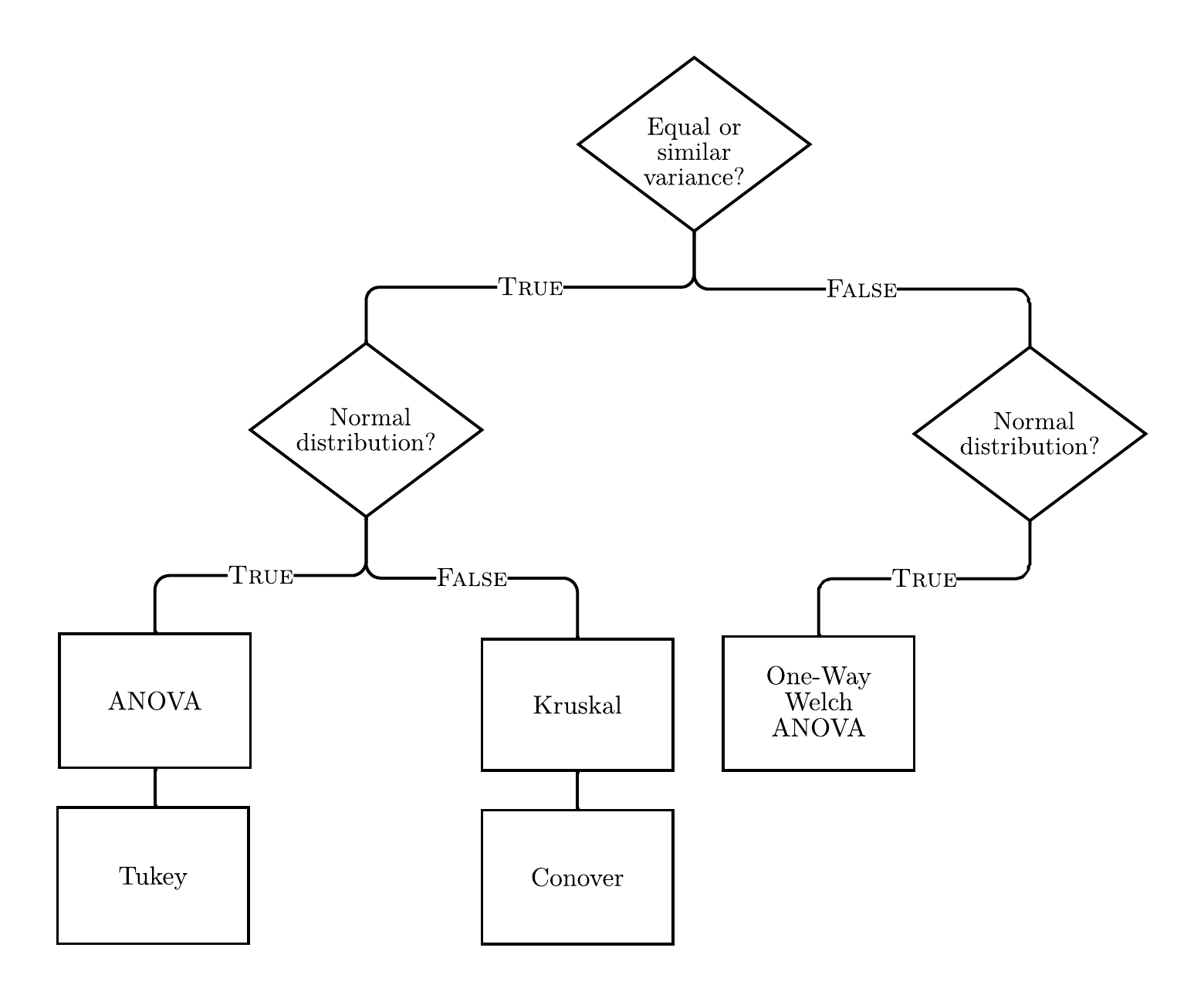}
    \caption{Flow diagram for hypothesis testing using AsaPy.}
    \label{fig:hypothesis-testing}
\end{figure}

One of the main features of the library is hypothesis testing, which determines whether a specific hypothesis about the data is true or false~\cite{kruschke2013bayesian}. This component offers a collection of statistical tests from which to choose. Focusing on the needs of the typical analyst, the module utilizes the decision flow depicted in Figure~\ref{fig:hypothesis-testing} to automatically select an appropriate test for the data being analyzed. For instance, AsaPy can streamline the process of conducting an ANOVA (Analysis of Variance) test, not only by performing the test itself but also by including all requisite pre-test checks and data visualizations to aid in interpreting the results. Integrating a full analytical pipeline -- from preliminary data checks to post-analysis interpretation tools -- represents a methodological advancement. This ensures that users are not only able to execute the desired statistical tests but also do so with a comprehensive understanding of the prerequisites and implications of these tests.

Another functionality is the distribution fit technique~\cite{taylor2017introduction}, used to adjust a particular statistical distribution to the data. This component gives information about the distribution that best fits the input data among the most common ones~\cite{gupta2018theory}, such as the normal, uniform, exponential, chi-squared, and beta distributions. The process of fitting distributions involves estimating their parameters, allowing us to extrapolate data beyond the range of the observed values. 

This package also includes methods for determining feature scores \cite{jolliffe2016principal}, which are used to rank the importance of different attributes in the dataset. These techniques are particularly helpful in filtering out non-informative or redundant variables from the input data~\cite{kuhn2013applied}. 

In addition, the package includes tools for Pareto front analysis~\cite{deb2002fast}, used to identify the optimal trade-off between two conflicting objectives. This is a useful tool for decision-making in complex systems. 

Furthermore, the analysis package also provides methods for detecting and removing outliers in the data~\cite{rousseeuw2019detecting}, which can significantly affect the accuracy of the analysis. This is achieved using various statistical techniques, such as the standard deviation and interquartile range methods.

Finally, the package provides exploratory data analysis (EDA) tools~\cite{tukey1977exploratory}, which can be used to visualize and understand the data. This toolkit consists of a table of class balance for categorical variables, association, correlation, histograms, and boxplots with information on the number of outliers for numerical variables. These preliminary analyses can help identify patterns, trends, and anomalies in the data and guide the choice of statistical models and analysis methods.

\subsection{Prediction}

The prediction step, the last one, uses the \texttt{models} module and provides a comprehensive framework for building custom machine learning models, including but not limited to neural networks and random forests. This package covers the entire process of creating a model, including phases such as data preprocessing, hyperparameter tuning, cross-validation, evaluation, and prediction. The most significant advantage of using this module is that it allows us to obtain estimated results without performing new simulations, thus saving time and computational resources. 

The package is built on top of popular machine learning libraries, such as TensorFlow~\cite{abadi2016tensorflow} and Scikit-learn~\cite{pedregosa2011scikit}. This allows for easy integration with other machine learning tools and workflows, aligned with the cardinal points conceived during the package design.

The \texttt{models} module provides various preprocessing methods to transform raw data into a format that can be used by machine learning algorithms. These methods include scaling, normalization, feature engineering, and more. It also provides options for handling missing values and categorical features, along with a wide range of available models. To help users effectively utilize these features, we provide some tutorials in our repository to guide the process of creating and evaluating models using the \texttt{models} package. 

Hyperparameters are parameters that are not learned during the training process but are set before the training begins. These hyperparameters can have a significant impact on the performance of the model. The \texttt{models} package includes methods for hyperparameter tuning, such as random search~\cite{bergstra2011algorithms}, which optimizes model performance by establishing the most effective hyperparameters before the training process. 

Cross-validation is a technique used to evaluate the performance of a model. The \texttt{models} package includes various methods for performing cross-validation, including k-fold cross-validation~\cite{arlot2010survey}.

The \texttt{models} package provides various evaluation metrics to measure the performance of a model to solve regression or classification problems. These metrics include accuracy, precision, recall, F1 score, mean squared error, and more. The package also provides options for visualizing the model's performance using plots. 



\subsection{Support module}

The AsaPy library provides users with a primary support module, namely \texttt{utils}. This module supply additional tools and utilities to users for handling data.


The \texttt{utils} module is a helpful tool for performing mathematical calculations and conversions in various fields, including geodesy, physics, and engineering. This component contains an assortment of constants, methods, and functions for converting distance and angle measurement units and changing coordinate systems. One of the key features of the \texttt{utils} module is its ability to convert between different units of distance and angle measurement, such as meters, kilometers, feet, miles, radians, and degrees. The package includes methods for executing these conversions with ease, making it simple to switch between units as necessary.

Another essential aspect of the \texttt{utils} module is its support for various coordinate systems. This module provides methods for converting between different coordinate systems, such as geodetic, geocentric, and Cartesian coordinate systems. This can be especially useful for applications in geodesy and geolocation, where precise positioning is critical. For example, the Geod class in the \texttt{utils} module provides methods for converting between geodetic and Cartesian coordinates and calculating distances and bearings between points on the Earth's surface. The ECEF (Earth-centered, Earth-fixed) class, on the other hand, offers methods for converting between ECEF and geodetic coordinates and calculating the distance between points in ECEF space. In summary, the \texttt{utils} module provides users with essential tools and utilities for handling data, performing mathematical calculations and conversions, and changing coordinate systems.

\section{CASE STUDY: BEYOND VISUAL RANGE AIR COMBAT SIMULATIONS}
\label{sec:use-cases}

In this section, we demonstrate how AsaPy can be effectively used to analyze military operational scenarios, especially in the context of beyond visual range (BVR) air combat simulations. BVR air combat is a challenging and critical field in which engagements occur beyond the pilot's visual range~\cite{dantas2018apoio}. These engagements make use of advanced weaponry and sensor systems. AsaPy has been employed in various applications to enhance BVR air combat simulations that feature agents controlled by artificial intelligence models~\cite{dantas2023autonomous}, demonstrating its capabilities and versatility. This section will discuss three primary applications of AsaPy, as implemented in other works using ASA and other simulation software.

Toward the end of the section, we introduce a new example analysis featuring a BVR fighter aircraft navigation scenario using ASA. This scenario serves as a practical illustration of AsaPy's applicability in a specific aerospace context, further emphasizing its role as a versatile tool within the aerospace domain.

\subsection{Engagement Decision Support}

The study conducted by \citet{dantas2021engagement} aimed to develop an engagement decision support tool for BVR air combat in the context of Defensive Counter Air (DCA) missions. In BVR air combat, engagement decision refers to the moment when the pilot decides to engage a target by executing corresponding offensive maneuvers. 

To plan the execution of simulations, the authors used the \texttt{doe} module from AsaPy, selecting key variables, including categorical and numerical with different coverage ranges, to simulate a BVR air combat. The simulation data was pre-processed and explored using the \texttt{analysis} module to organize and better understand the data. These variables included the distance, angle between the longitudinal axis, and difference in altitude between the reference and the target. The authors ran 3,729 constructive simulations that lasted 12 minutes each, resulting in 10,316 engagements. 

The authors evaluated the simulations using an operational metric called the DCA index, which represents the degree of success in this type of mission based on the expertise of subject matter experts. The DCA index is based on the distances between aircraft from both the same team and opposing teams, as well as the number of missiles deployed. The index indicates the likelihood of success in BVR air combat during DCA missions. The primary aim of these missions is to establish a Combat Air Patrol, which requires aircraft to fly in a specific pattern around a designated location. 

The authors employed the \texttt{models} module from AsaPy to build a supervised machine learning model based on decision trees to determine the quality of a new engagement, using the engagement status right before it starts and the average of the DCA index throughout the engagement. Beyond model creation, the authors seamlessly integrated AsaPy for data preprocessing and hyperparameter tuning as well. Overall, the authors utilized various features of the AsaPy library to plan, execute, and analyze their simulations and to build and evaluate a model for engagement decision support in BVR air combat.


\subsection{Weapon Engagement Zone Evaluation}

Still in the BVR air combat context, \citet{dantas2021weapon} used AsaPy to analyze simulation data generated by the ASA simulation environment, explicitly focusing on calculating an air-to-air missile's weapon engagement zone (WEZ). The WEZ allows the pilot to identify airspace where the available missile is more likely to successfully engage a particular target, i.e., a hypothetical area surrounding an aircraft where an adversary is vulnerable to a shot.


Designing experiments for missile launches in BVR air combat is a complex process that involves considering various input variables. These variables help to simulate different scenarios and identify the best possible outcomes for missile launches. The seven input variables for the simulation runs include the shooter altitude, shooter speed, target altitude, target speed, target heading, the relative position of the target, and shooter pitch. 

Each variable plays a crucial role in determining the WEZ maximum range. The authors used the \texttt{doe} module to perform the LHS method from AsaPy to plan and design the simulation experiments using these variables.

The simulations were executed in chunks to improve the computational efficiency and reduce the simulation time. After completing the simulation runs, the authors collected the output data and analyzed it using the \texttt{analysis} package in AsaPy, generating data visualization, feature engineering, and statistical tests to understand the data distribution and identify relationships between the input variables and the WEZ. 

Using the data from the simulations and analysis, the authors built a supervised machine learning model using a Deep Neural Network (DNN) to predict the WEZ maximum launch range for a given scenario.

Finally, the authors used the \texttt{models} package in AsaPy to build and train the model to predict the WEZ maximum launch range for a given scenario. They evaluated its performance using metrics such as the mean absolute error and the coefficient of determination to ensure that the model accurately predicted the WEZ for different scenarios.

In a related study presented by \citet{dantas2023real}, the emphasis was on the WEZ of Surface-to-Air Missiles (SAMs). SAMs hold a crucial position in the landscape of modern air defense systems. The WEZ's significance is further accentuated as it is directly associated with a missile’s maximum range, marking the farthest interception distance between a missile and its target.

Conventional simulation methods, in many instances, result in significant computational demands and extended processing times. To address these challenges, the study incorporated machine learning techniques, using AsaPy prediction methods synergized with specialized simulation tools to train supervised algorithms. Through AsaPy, researchers were able to streamline the simulation and analysis process, making it more efficient and data-driven.

By utilizing a comprehensive dataset from earlier SAM simulations, the model demonstrated remarkable accuracy in predicting the SAM WEZ based on new input parameters. This combination of machine learning and advanced simulation tools not only accelerated SAM WEZ simulations but also bolstered strategic planning in air defense, offering invaluable real-time insights that enhance the performance of SAM systems. The study, through AsaPy, provided an in-depth analysis of different machine learning algorithms, elucidating their capabilities and performance metrics. It not only suggested avenues for future research but underscored the transformative potential of incorporating machine learning into SAM WEZ simulations.

\subsection{Missile Hit-Prediction}

In~\citet{dantas2022machine}, the authors analyzed both defensive and offensive scenarios in BVR air combat using AsaPy. The authors used AsaPy to generate constructive simulations of BVR air combat scenarios and extract various features related to situational awareness from the simulation data. They designed a multilayer perceptron neural network incorporating data from these simulations to enhance pilots' situational awareness during in-flight decision-making. By training their machine learning models based on neural networks on this data, they could accurately predict a pilot's situational awareness based on the missile's ability to hit the target. Therefore, \citet{dantas2022machine} demonstrated the potential of machine learning in BVR air combat scenarios by generating fast and reliable responses concerning the tactical state to improve the pilot's situational awareness and, therefore, the in-flight decision-making process.

\begin{figure*}[ht]
\centering
    \includegraphics[width=0.89\textwidth]{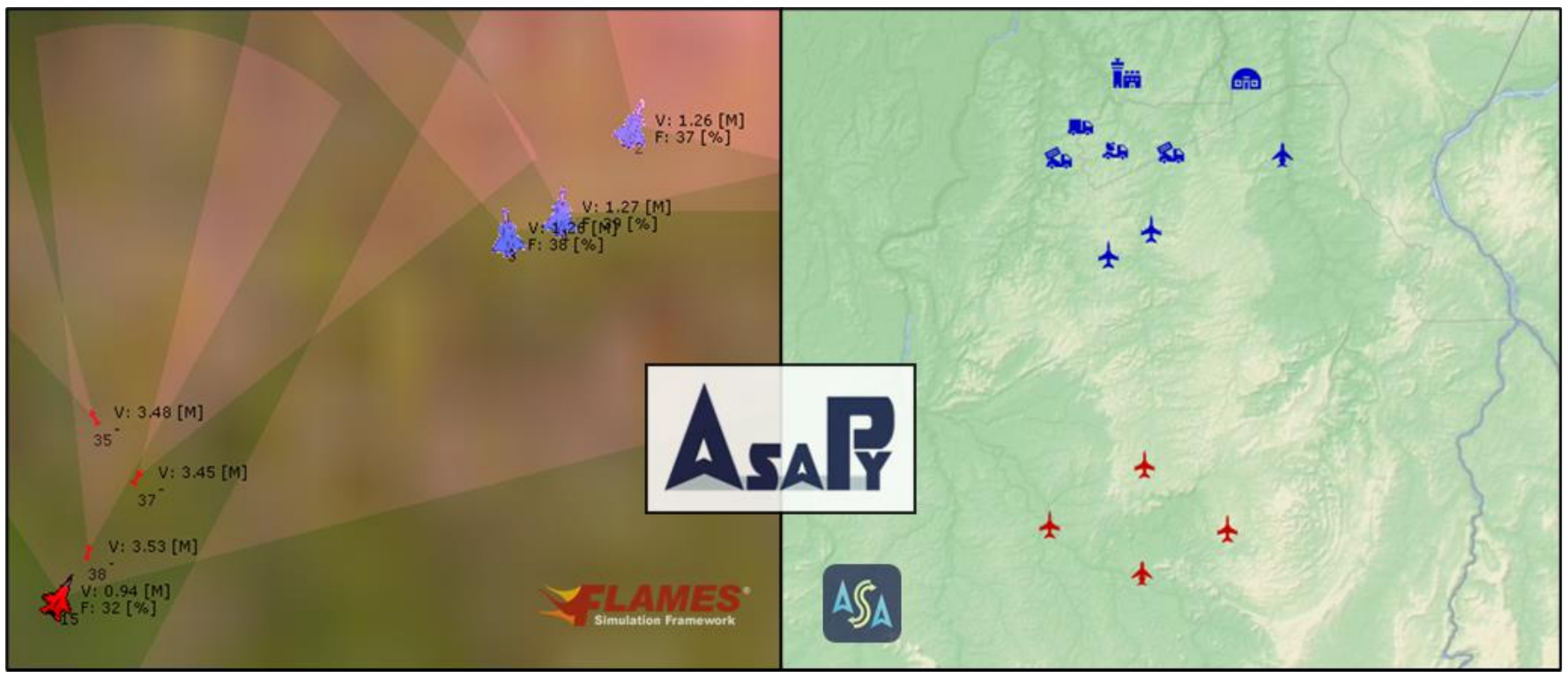}
    \caption{Examples of simulation platforms in which AsaPy may be employed: FLAMES (left) and ASA (right).}
    \label{fig:asapy-cases}
\end{figure*}

One of the key strengths of AsaPy is its ability to work with various simulation software, including commercial and open-source platforms (Figure~\ref{fig:asapy-cases}). This versatility allows users to leverage the power of AsaPy regardless of the simulation software they are using, making it a valuable tool for military operational scenario analysis. For example, in~\citet{dantas2022supervised}, the authors aim to develop a machine learning model that can predict the effectiveness of missile launches in BVR air combat scenarios. To generate the simulation data, the authors used the FLAMES simulation platform, a commercially available simulation software suite. The AsaPy library was used to organize and analyze the simulation data generated by FLAMES. The authors used AsaPy to build seven different supervised machine learning models that predict the effectiveness of missile launches in BVR air combat scenarios. To improve the performance of the machine learning models, the authors also used resampling techniques such as SMOTE~\cite{chawla2002smote} to generate more data on missile launches. This approach helped to address the class imbalance issue that commonly occurs in military operational scenarios, where successful missile launches are relatively rare compared to unsuccessful ones.

Overall, the successful implementation of AsaPy in those missile hit-prediction works demonstrates its versatility and utility in analyzing and modeling military operational scenarios. The AsaPy library proved a valuable asset in the data analysis and modeling process for the aforementioned works, providing efficient data preprocessing and analysis tools and aiding in developing models to solve complex problems in the BVR air combat context. AsaPy's compatibility with various simulation software highlights its versatility, making it an excellent choice for researchers and practitioners in the military and defense sectors who frequently work with multiple simulation platforms.

\subsection{Fighter Aircraft Navigation}

In this subsection, we explore the complexities of fighter aircraft navigation, examining the interconnections between various flight parameters and their impact on fuel efficiency. This analysis, supported by detailed experimental data, aims to deepen our understanding of efficient aircraft operation.

The scenario under examination involves a navigation flight executed by a fighter aircraft, characterized by diverse maneuvers at various altitudes and speeds. The aircraft navigates between 10,000 and 35,000 feet, adjusting its speed from 350 to 550 knots. Additionally, the flight includes a 10-minute holding maneuver at the third route point, where the aircraft follows a circular path in the air. This maneuver is typically used for traffic management or to delay the aircraft before landing.

This specific case study comprises two experiments aimed at elucidating the methodology of extracting and analyzing data from simulations to address pertinent questions in aerospace studies. For an in-depth understanding of this process, we invite you to examine the code associated with our analysis.

\subsubsection{Experiment 1 -- Analysis of the Relationship between Time of Flight and Fuel Consumption:} 

The first experiment investigates the link between time of flight, denoted as \texttt{time\_of\_flight}, and fuel consumption, referred to as \texttt{fuel\_consumed}, in a flight simulation scenario. The main goal is determining if longer flight durations directly relate to increased fuel consumption. This involves analyzing data from 4,000 flight simulations, focusing on total flight duration in seconds and the amount of fuel consumed in pounds.

\begin{figure}[ht]
\centering
    \includegraphics[width=0.485\textwidth, trim={10pt 12pt 0pt 25pt}, clip]{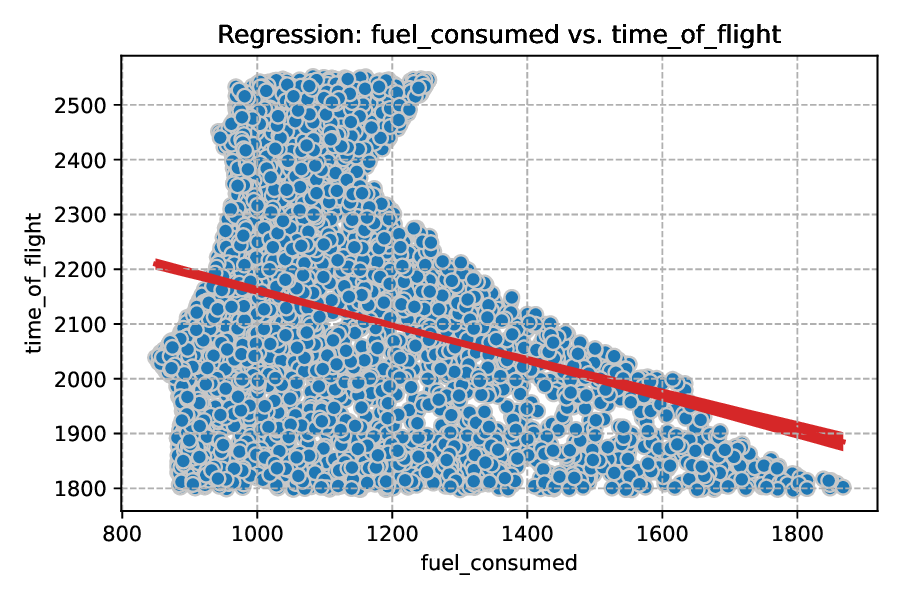}
  \caption{Linear regression of Time of Flight vs. Fuel Consumed.}
    \label{fig:linear}
\end{figure}

To accomplish this, various statistical methods, including linear regression and correlation analysis, are used. Furthermore, data visualization techniques, especially scatter plots, are utilized to interpret the relationship between these key variables (Figure~\ref{fig:linear}).

The central theory suggests a direct, positive correlation between time of flight and fuel consumption, indicating that longer flights generally result in higher fuel usage. However, unexpectedly, the results show no clear linear relationship between these variables. This surprising finding is mainly due to variations in speed and altitude during the simulations, indicating that these factors significantly affect fuel consumption dynamics. This complexity, going beyond simple linear correlations, highlights the need for more research into how speed and altitude change influence fuel consumption.

\subsubsection{Experiment 2 -- Analysis of the Relationship between Speed, Altitude, and Fuel Consumption:}
\begin{figure*}[ht]
    \centering
    \begin{subfigure}[b]{0.485\textwidth}
        \includegraphics[width=\textwidth, trim={50pt 30pt 120pt 65pt}, clip]{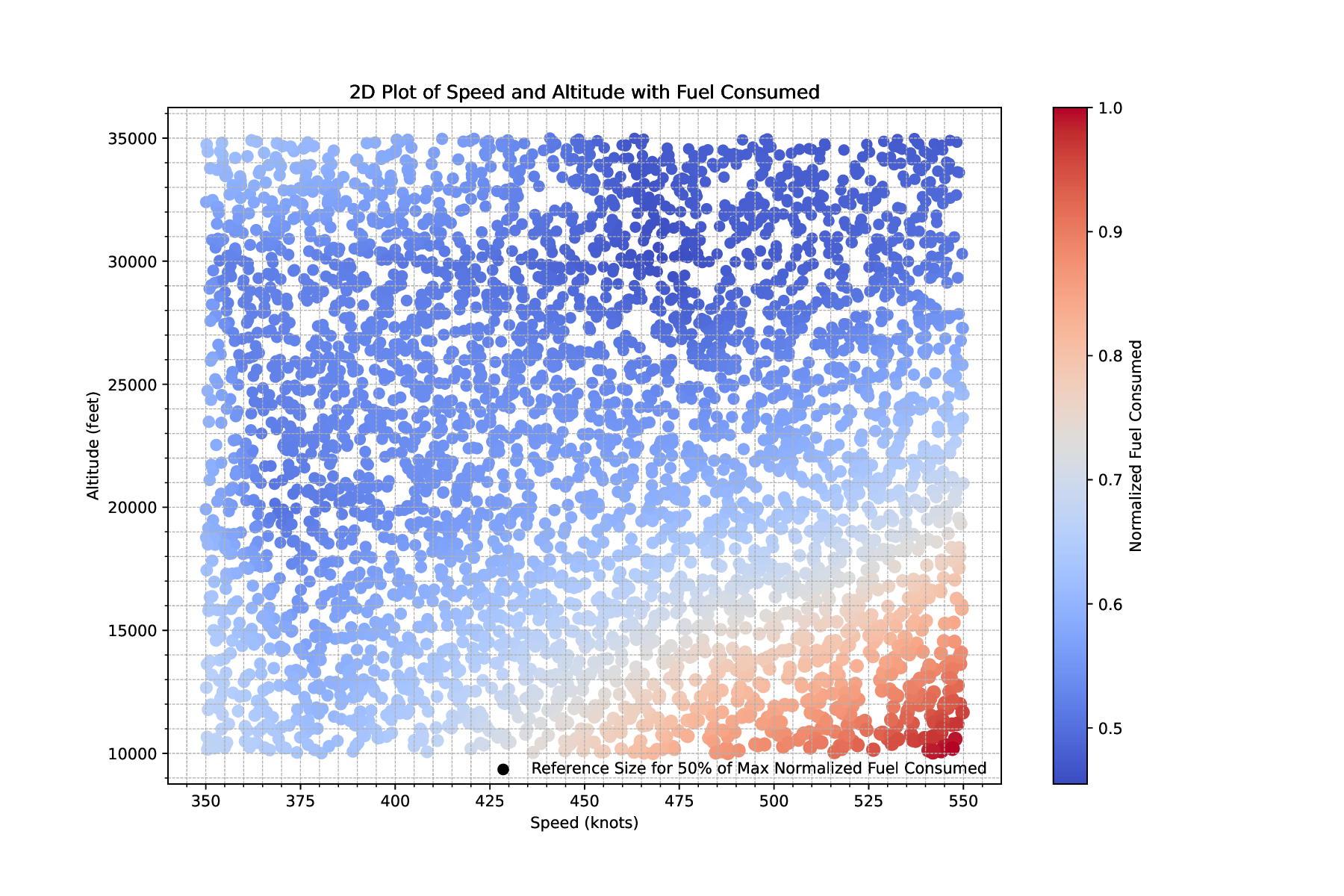}
        \caption{2D normalized surface plot showing the relationship between speed, altitude, and fuel consumption.}
        \label{fig:2d}
    \end{subfigure}
    \hfill
    \begin{subfigure}[b]{0.485\textwidth}
        \includegraphics[width=\textwidth, trim={200pt 130pt 160pt 150pt}, clip]{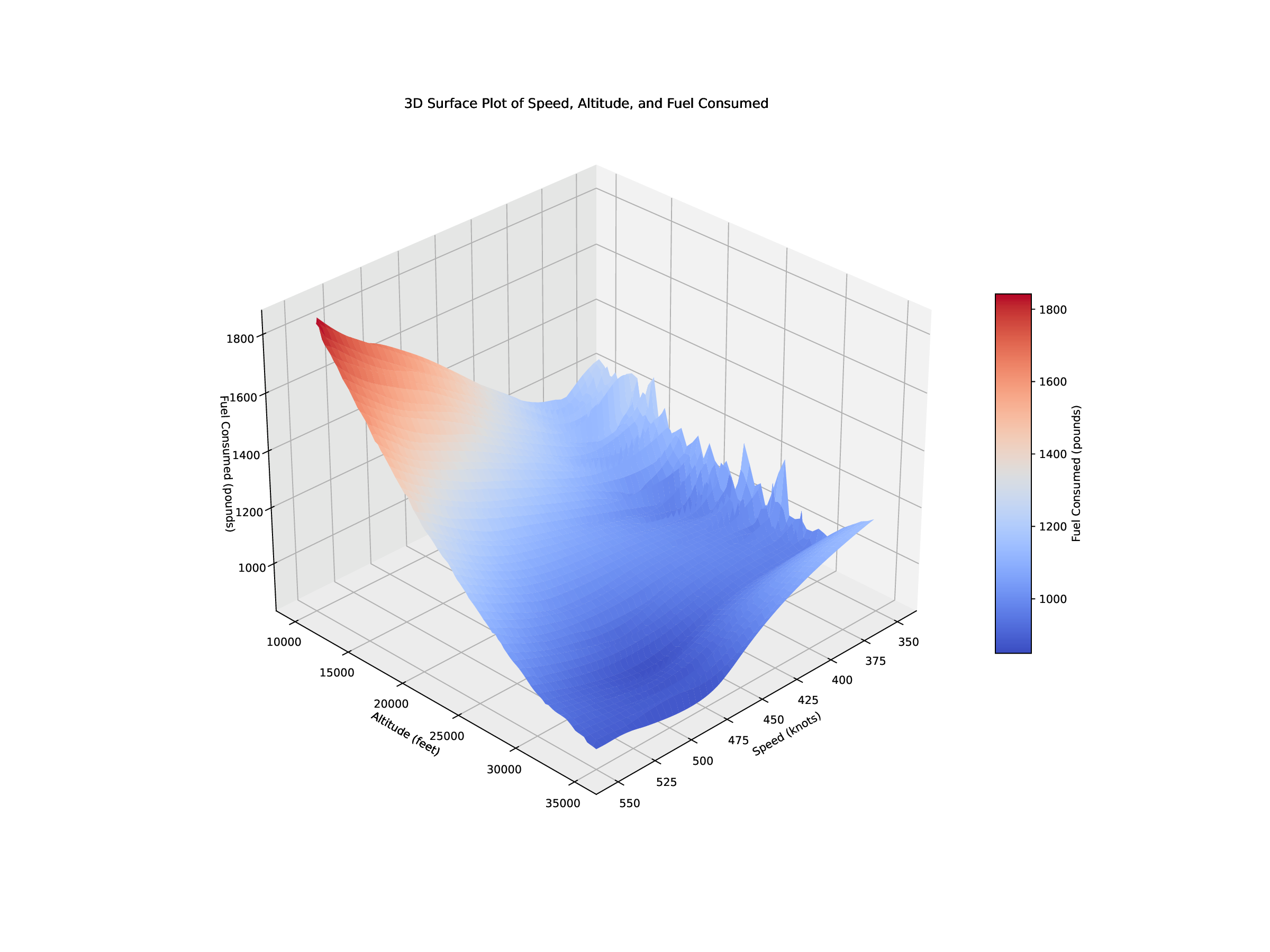}
        \caption{3D surface plot illustrating the dynamic interaction between speed, altitude, and fuel consumption.}
        \label{fig:3d}
    \end{subfigure}
    \caption{Comparative analysis of aircraft performance parameters. (a) presents a 2D perspective, while (b) offers a 3D visualization, providing a comprehensive overview of fuel efficiency dynamics.}
    \label{fig:2d-3d-combined}
\end{figure*}

Expanding on the initial experiment's results, the second experiment clarifies the relationship between flight speed, altitude, and fuel consumption in a simulation context. The goal is to understand how these factors, both individually and together, impact an aircraft's fuel efficiency. This investigation involves analyzing extensive flight simulation data, focusing on the interplay between speed in knots, altitude in feet, and fuel consumption measured in pounds. The study reveals complex relationships and patterns using statistical analyses, two-dimensional charts, and surface plots (Figure~\ref{fig:2d-3d-combined}).

The primary hypothesis of the experiment suggests that both speed and altitude significantly affect fuel consumption in a potentially complex and interactive manner. The research aims to determine if higher speeds or altitudes proportionally increase fuel consumption and to identify optimal efficiency points.

One of the key findings of the experiment is the discovery of fuel consumption peaks. These peaks are most prominent in higher and red regions of the chart, occurring at an altitude of approximately 10,000 feet and a speed of about 525 knots. At these parameters, the consumption reaches nearly 1800 pounds. This insight helps to understand the conditions under which fuel consumption is maximized, informing operational and design decisions for aircraft.

In contrast, the study also identifies areas of operational efficiency. These are indicated by blue areas on the chart, representing lower fuel consumption and suggesting more efficient operating ranges. The lowest consumption values, close to 1000 pounds, are observed at altitudes of around 25,000 to 30,000 feet and speeds between 400 to 450 knots. This finding is significant for optimizing flight paths and aircraft design for energy efficiency.

Additionally, the experiment reveals a complex variation in fuel consumption at certain intermediate speeds and altitudes. This observation indicates an operational efficiency point that does not follow a simple linear relationship with speed or altitude, adding a layer of complexity to the understanding of aircraft fuel efficiency.

Moreover, the insights gained from the chart are invaluable for planning routes that prioritize fuel efficiency. By avoiding altitude and speed ranges that result in excessive consumption, significant improvements in operational cost can be achieved.

Finally, the experiment's results reveal insights into aircraft performance. The data illustrate how the engine and aircraft perform under various operational conditions, aiding engineers in optimizing or developing more efficient propulsion systems. These findings are important for advancing the field of aeronautical engineering and contribute to the development of more efficient aircraft.

\section{CONCLUSION AND FUTURE WORK}
\label{sec:conclusions}
\balance
In conclusion, AsaPy distinguishes itself as a versatile Python library that streamlines and accelerates the analysis of simulation data. Its features, including experiment design, statistical analysis, machine learning algorithms, and data visualization tools, make this library a good resource for engineers and researchers in simulation studies, particularly in the aerospace and military domains.

Future work on AsaPy is multifaceted, aiming at both enhancement and expansion. We plan to integrate additional DoE methods and machine learning algorithms to broaden its applicability. Another priority is optimizing the performance of AsaPy's algorithms, to enable faster processing and the handling of larger datasets. Enhancing interoperability through integration with other tools and platforms is also on our agenda, further improving usability. Continual refinement of the documentation and user interface will make AsaPy more user-friendly and accessible.

A key focus of our future work is the practical demonstration of AsaPy's effectiveness in real-world scenarios. We propose a comprehensive analysis of AsaPy's impact on data analysis efficiency. This would involve contrasting the processes of managing simulation output data from different systems, such as FLAMES or ASA, with and without AsaPy. The emphasis will be on assessing how AsaPy streamlines tasks like data reading, loading, cleaning, and preliminary analyses.

Additionally, we are planning to expand our suite of data analysis tools, particularly focusing on expanding supervised learning algorithms and introducing unsupervised learning methods, such as clustering and principal component analysis (PCA). This expansion aims to enhance AsaPy's ability to uncover patterns and relationships in data without the need for pre-labeled outcomes.

\begin{acks}
This work was supported by Finep (Reference No. 2824/20). Takashi Yoneyama and Marcos R. O. A. Maximo received partial funding from CNPq – National Research Council of Brazil, under grants 304134/2-18-0 and 307525/2022-8, respectively.
\end{acks}

\section*{Source Code}
AsaPy is available as an open-source library. You can download it from \href{https://github.com/ASA-Simulation/asapy}{https://github.com/ASA-Simulation/asapy}.

\bibliographystyle{ACM-Reference-Format}
\bibliography{ref}

\end{document}